\documentclass[twocolumn]{jpsj2} 
\title{$^{17}$O NMR Measurements on Superconducting Na$_{0.35}$CoO$_{2} \cdot y$H$_{2}$O}
\author{
Yoshihiko \textsc{Ihara,}\thanks{E-mail address: ihara@scphys.kyoto-u.ac.jp}
Kenji \textsc{Ishida,}\thanks{E-mail address: kishida@scphys.kyoto-u.ac.jp}
Kazuyoshi \textsc{Yoshimura,}$^{1}$
Kazunori \textsc{Takada,}$^{2}$
Takayoshi \textsc{Sasaki,}$^{2}$
Hiroya \textsc{Sakurai,}$^{3}$
and
Eiji \textsc{Takayama-Muromachi}$^{3}$}

\inst{
Department of Physics, Graduate School of Science, Kyoto University, Kyoto 606-8502, Japan \\
$^{1}$Department of Chemistry, Graduate School of Science, Kyoto University, Kyoto 606-8502, Japan \\
$^{2}$Advanced Materials Laboratory, National Institute for Materials Science, 1-1 Namiki, Tsukuba, Ibaraki 305-0044, Japan\\
$^{3}$Superconducting Materials Center, National Institute for Materials Science, 1-1 Namiki, Tsukuba, Ibaraki 305-0044, Japan
}

\abst{
An $^{17}$O NMR measurement was performed on nonoriented polycrystalline Na$_{0.35}$CoO$_2 \cdot y$H$_2$O with superconducting (SC) transition temperature $T_\text{c} = 4.6$ K. 
A weak temperature dependence was observed in the Knight shift at the O site ($^{17}K$). The spin part of $^{17}K$ ($^{17}K_{\rm spin}$) is estimated from the plot of $^{17}K$ against bulk susceptibility $\chi$. 
The $^{17}K_{\rm spin}$ decreases in the SC state, indicative of the decrease in the in-plane component of the spin susceptibility. 
The nuclear spin-lattice relaxation rate $1/T_1$ at the O site $^{17}(1/T_1)$ shows a good scaling with $1/T_1$ at the Co site $^{59}(1/T_1)$. 
This indicates that the spin fluctuations at the O site originate from the Co spin dynamics. 
The relationships between $^{17}(1/T_1T)$ and $^{17}K_{\rm spin}$ and between $^{17}(1/T_1T)$ and $^{59}(1/T_1T)$ show the development of incommensurate fluctuations at {\bf q $\sim$ 0} other than {\bf q $=$ 0} below 30 K. 
A clear indication of ferromagnetic correlations at {\bf q $=$ 0} was not observed from the present $^{17}$O-NMR studies.}

\kword{superconductivity, hydrate sodium cobalt oxide, NMR, spin fluctuations}

\begin{document}
\maketitle

The recently discovered hydrate cobaltate superconductor Na$_x$CoO$_2\cdot y$H$_2$O has attracted much attention\cite{takada}, because its two-dimensional CoO$_2$ layers where superconductivity occurs form a triangular structure, in contrast to the tetragonal structure in cuprate superconductors. An interplay between superconductivity and geometrical frustrations is invoked due to the crystal structure. Various experiments as well as the spin-lattice relaxation rate $1/T_1$ in the superconducting (SC) state revealed that this superconductivity is of an unconventional type\cite{ishida,fujimoto,yang}, where electron correlations play an important role in superconductivity. 
So far, we have shown the relationship between magnetic fluctuations and SC transition temperature $T_{\text{c}}$ from the Co nuclear-quadrupole-resonance (NQR) measurements on various samples with different values of $T_{\text{c}}$ \cite{ihara1,ihara2}: $T_{\text{c}}$ increases with increasing spin fluctuations at $T_{\text{c}}$, and the highest $T_{\text{c}}$ in the system is observed in the vicinity of the magnetic phase.\cite{ihara2} In addition, it has been shown that the spin fluctuations correlate with NQR frequency $\nu_{\text{Q}}$, which is related to the distortion of the CoO$_6$ octahedron along the $c$-axis.\cite{ihara2} Therefore, we suggest from an experimental point of view that the crystal-field splitting between the $a_{1g}$ and $e_{g'}$ states of Co-$3d$ $t_{2g}$ orbitals is an important parameter for determining superconductivity. \cite{ihara2}  Quite recently, a scenario based on experimental results has been proposed from a theoretical point of view \cite{mochizuki}. To verify this scenario, it is quite important to identify the properties of the magnetic fluctuations in the normal state and the symmetry of the SC pairs. Spin-triplet superconductivity induced by ferromagnetic fluctuations is expected in several theoretical models \cite{Ikeda,yanase,mochizuki}. 

To date, there have been several reports about the $^{59}$Co Knight shift ($^{59}K$) in the SC state.\cite{kobayashi1,kobayashi2,waki} Although a decrease in $^{59}K$ was observed in the SC state, quantitative discussion was not carried out due to an ambiguity in the spin part of $^{59}K$.\cite{kobayashi1,kobayashi2} This ambiguity originates mainly from the large electric field gradient (EFG) with a temperature dependence and the large orbital part of $^{59}K$. Therefore, we change the NMR nucleus to $^{17}$O, since $^{17}$O has a small EFG frequency ($\nu_Q$) and a small orbital shift in general.

In this letter, we report on the spin part of the Knight shift at the O site ($^{17}K_{\rm spin}$) estimated from a plot of $^{17}K$ against $\chi$, and the behavior of $^{17}K_{\rm spin}$ in the SC state. A decrease in $^{17}K$ is observed in the field up to 8 T, which is consistent with the results in $^{59}K$ \cite{kobayashi1,kobayashi2}. In addition, we found the development of AFM fluctuations below 30 K from the measurement of $1/T_1$ at the O site. Plausible spin-fluctuation character is discussed on the basis of our NMR experiments.

The sample we used in the measurement shows superconductivity at $T_{\text{c}} \sim 4.6$ K, which was determined by a dc susceptibility measurement.
The resonance frequency of the $\pm 5/2 \leftrightarrow \pm7/2$ transition of Co NQR is 12.45 MHz, slightly larger than that of the highest-$T_c$ sample (12.40 MHz).\cite{ihara2} According to the phase diagram we developed,\cite{ihara2} this sample is situated near the border between superconductivity and magnetism. Strong magnetic fluctuations with critical character are expected in this sample.  
The detailed sample preparation and the $^{17}$O exchange procedures will be reported elsewhere.\cite{sakurai1}
$^{16}$O nuclei in the CoO$_{2}$ layers are partially replaced by $^{17}$O isotopes with the nuclear spin $I = 5/2$, but O of the H$_{2}$O are not. This is because the exchange annealing was carried out in anhydrate Na$_{0.7}$CoO$_2$, and then water was added to the compound. An $^{17}$O NMR measurement was carried out using the nonoriented powder sample because mixing some material to fix the powdered sample orientation might degrade sample quality.

\begin{figure}[htbp]
\begin{center}
\includegraphics[width=7cm,clip]{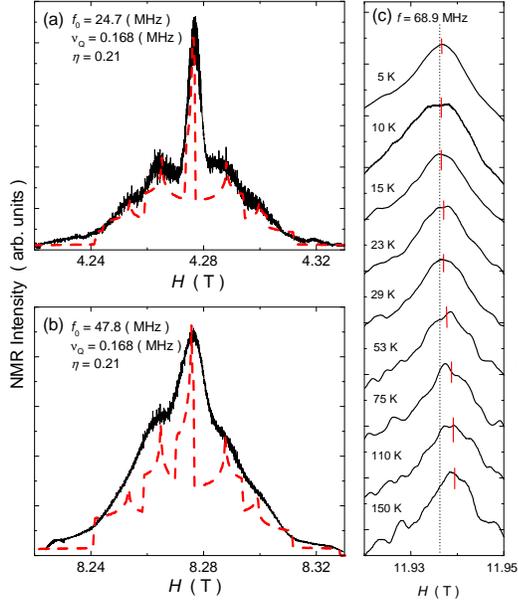}
\caption{NMR spectra obtained for the powder sample at the frequencies of (a) 24.7 and (b) 47.8 MHz. 
The dashed lines are calculated powder patterns. The powder-pattern spectra are well fitted by the calculation with $\nu_{Q} =0.168$ MHz and $\eta=0.21$ (see in text). (c) The central peak of the NMR spectrum recorded in the frequency of 68.9 MHz at various temperatures.  }
\label{spectra}
\end{center}
\end{figure}

Figures 1 (a) and 1 (b) show NMR spectra taken at the frequencies of 24.7 and 47.8 MHz. The spectra show the typical powder-pattern structure with a non-zero asymmetric parameter $\eta= (\nu_{xx}-\nu_{yy})/\nu_{zz}$ ($\nu_{\alpha \alpha}$ : EFG along the $\alpha$ direction).   
In general, when nuclei with a nuclear quadrupole moment are in a magnetic field, the total interaction is a sum of the Zeeman and electric-quadrupole (eqQ) interactions and is written as \cite{MetallicShift}
\begin{align}
\mathcal{H} &=\mathcal{H_{\rm Zeeman}} + \mathcal{H_{\rm eqQ}} \notag \\ 
           &=(1+K)\gamma_n\hbar I\cdot H + \frac{\hbar\nu_\text{Q}}{6}\Big\{ (3I_{z}^{2}-I^{2})+\frac{1}{2}\eta(I_{+}^{2}+I_{-}^{2})\Big\}. \notag
\end{align}
Here, $K$ and $\gamma_n$ are the Knight shift and nuclear gyromagnetic ratio, respectively. When the Zeeman interaction is dominant, $\mathcal{H_{\rm eqQ}}$ is treated as a perturbation.
The eigenvalue of the total Hamiltonian depends on the angle between the applied field and the principal axis of EFG.
In a nonoriented powder sample, the principal axis of the EFG is randomly distributed in all directions with respect to the external field.
The powder-pattern spectrum observed for the nonoriented sample is reproduced using fitting parameters of center field $H_{0}$, $\nu_\text{Q}$, $\eta$ and $K$.
Here, $\nu_{\rm Q}$ and $\eta$ are evaluated to be 0.168 MHz and 0.21, respectively, both of which are independent of applied field. 
We note that the $\eta$ at the O site is close to that at the Co site (0.208 $\pm$ 0.007) \cite{fujimoto,ihara2}.

The powder-pattern spectrum of the $^{17}$O-NMR becomes broader with increasing field, resulting in a structureless spectrum as seen in Fig.~1 (b). This is due to the large susceptibility at low temperatures. \cite{sakurai2} Although the $^{17}$O-NMR spectrum becomes broader in higher magnetic fields, the Knight-shift measurement in high fields is effective for detecting the temperature dependence up to higher temperatures because the difference between the resonance field and the reference one becomes larger with increasing applied field. Fig. 1 (c) shows the central peak in the NMR spectrum recorded at various temperatures at the frequency of 68.9 MHz.

\begin{figure}[htbp]
\begin{center}
\includegraphics[width=7.5cm,clip]{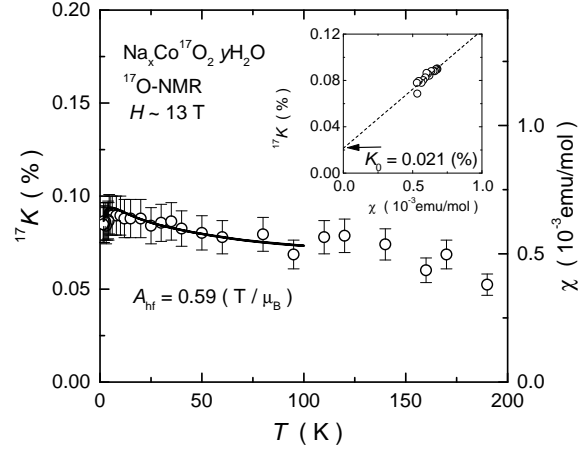}
\caption{Temperature dependences of $^{17}$O Knight shift $^{17}K$ (left axis) measured in magnetic field of 13 T and dc susceptibility in one of 7 T (right axis). 
$\chi$ was corrected for a diamagnetic effect from the core electrons of -7.4 $\times 10^{-5}$ emu/mol. The inset shows a plot of $^{17}K$ against $\chi$. 
From the linear relationship between $^{17}K$ and $\chi$, the hyperfine coupling constant and the temperature-independent $^{17}K_0$ are estimated to be 0.59 T/$\mu_{\text{B}}$ and 0.021$\pm 0.006$\%, respectively.}
\label{Kchi}
\end{center}
\end{figure}
The temperature dependence of the Knight shift of $^{17}$O ($^{17}K$) was measured in a magnetic field of 13 T, and is shown in Fig. \ref{Kchi}.
Although the typical experimental error is approximately 0.02\% due to the broad NMR peak, the Knight shift shows a temperature dependence beyond the error bars as shown in Fig.~\ref{Kchi}. 
The $^{17}K$ gradually increases with decreasing temperature, which is scaled with $\chi$ measured in a magnetic field of 7 T\@. 
The inset of Fig. \ref{Kchi} shows the plot of $^{17}K$ against $\chi$ with temperature as an implicit parameter.
A good linear relationship was observed between $\chi$ and $K$, showing that the weak temperature dependence of bulk susceptibility is an intrinsic effect. 
From the linear relationship, the hyperfine coupling constant $^{17}A_{\rm hf}$ and temperature-independent Knight shift $^{17}K_0$ are evaluated to be 0.59 T/$\mu_{\rm B}$ and 0.021$\pm 0.006$\%, respectively.
Such a weak temperature dependence of $^{17}K$ would be difficult to detect in small magnetic fields.
$^{17}K_{0}$ is ascribed to the orbital shift and the shift as being due to the eqQ interaction. The spin part of $^{17}K$ ($^{17}K_{\rm spin}$) is estimated by subtracting $^{17}K_{0}$ from the total observed Knight shift.
It is shown that the orbital susceptibility $\chi_{\rm orb}$ is negligibly small and that the spin susceptibility $\chi_{\rm spin}$ is dominant in the total susceptibility.
Using the estimated $\chi_{\rm spin}$ and the Sommerfeld term ($\gamma_{\rm el} =$ 15 mJ/mol K$^{2}$) obtained in the specific-heat measurement reported previously\cite{yang}, the Wilson ratio $R_{\text{W}}$ is estimated to be $\sim$ 2.6.
The Wilson ratio greater than two suggests the enhancement of the spin susceptibility compared with that estimated from the specific-heat measurement.

\begin{figure}[htbp]
\begin{center}
\includegraphics[width=7.5cm,clip]{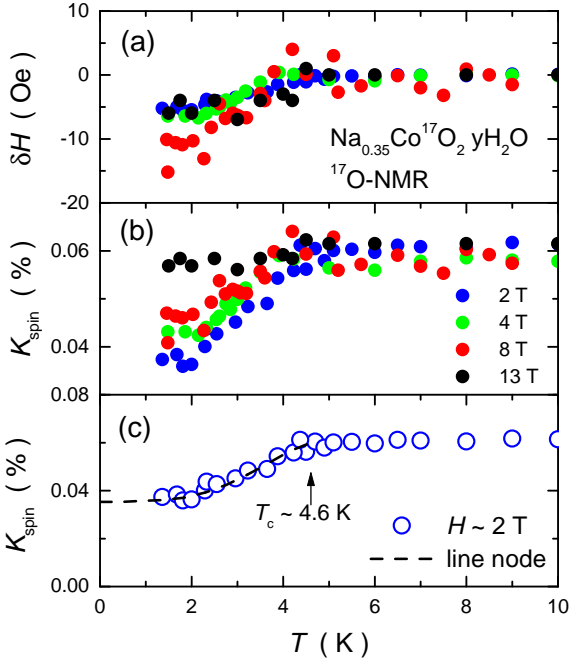}
\caption{Temperature dependence of (a) the relative shift of the magnetic field $\delta H$ with respect to the resonance field at $T = T_{\rm c}$, (b) $^{17}K_{\rm spin}$ in various fields around $T_{\rm c}$. (c) Quantitative analyses of Knight shift below $T_{\text{c}}$ using the two-dimensional line-node model (see in text).}
\label{KTc}
\end{center}
\end{figure}

Next, we discuss the variation of $^{17}K_{\rm spin}$ in the SC state. The resonance field of the central peak is shifted abruptly below $T_{\rm c}$. Figure \ref{KTc}(a) shows the relative shift of the central peak $\delta H$ below $T_{\rm c}$ with respect to the resonance field at $T_{\rm c}$.
In general, $\delta H$ in the SC state is ascribed to the SC diamagnetic effect $\varDelta H_{\rm dia}$, and the decrease in spin susceptibility is  given by $\varDelta H_{\rm spin} = \varDelta K_{\rm spin} \times H_0$, where $H_0$ is the applied field. These two effects exhibit different field dependences. $\varDelta H_{\rm dia}$ becomes smaller, whereas $\varDelta H_{\rm spin}$ becomes larger with increasing $H_0$.
The increase in $|\delta H|$ when the field increases from 4 to 8 T suggests that the effect of $\varDelta H_{\rm spin}$ is dominant in this field range.
Taking $^{17}K_{\rm spin} \sim$ 0.06\% into consideration, we discuss quantitatively the temperature variation  of $^{17}K_{\rm spin}$ in the SC state.
Figure \ref{KTc}(b) shows the temperature variation of $^{17}K_{\rm spin}$.
Since $^{17}K_{\rm spin}$ is determined from the central peak of the powder-pattern spectra, $^{17}K_{\rm spin}$ corresponds to the isotropic term of $^{17}K$, ($^{17}K_{\rm iso}=(^{17}K_a+^{17}K_b+^{17}K_c)/3$). 
According to the $H_{c2}$ versus $T$ phase diagram \cite{chou}, superconductivity is almost destroyed when $H$ = 2 T is applied along the $c$ axis. 
Thus, we consider that $^{17}K_c$ is unchanged, and that $^{17}K_a$ and $^{17}K_b$ decrease in the SC state.
The decrease in $^{17}K_{\rm spin}$ shows that the in-plane component of the spin susceptibility decreases in the SC state. 
This suggests that the superconductivity is in a spin-singlet state ($d$-wave in this case), or in a spin-triplet state with the SC {\bf d}-vector in the CoO$_2$ plane, 
since the spin susceptibility along the SC {\bf d}-vector decreases when magnetic fields are applied parallel to the {\bf d}-vector. 
The possibility of the spin-triplet state with the {\bf d}-vector along the $c$-axis is excluded since the spin susceptibility should be unchanged in the fields perpendicular to the {\bf d}-vector. 
The experimentally observed decrease in $^{17}K_{\rm iso}$ is reproduced by the two-dimensional line-node ($\Delta(\phi)=\cos(2\phi)$) model with the singlet pairing, 
which incorporates the residual density of states DOS ($N_{\rm res}$) originating from the imperfection and/or inhomogeneity of the compound\cite{ishida2}. 
The fitting parameters are $2\Delta/k_{\rm B}T_{\rm c} = 3.5$ and $N_{\rm res}/N_0 =0.32$ ($N_0$: DOS at the Fermi level), which were determined to reproduce the temperature dependence of $1/T_1$ in the SC state \cite{ishida}. It should be noted that the onset temperature of the decrease in $^{17}K$ is unchanged with respect to the applied field, which is in good agreement with the dc susceptibility measurement\cite{sakurai2}. It is likely that the insensitivity of $T_{\rm c}$ with respect to the strength of applied fields is related to the two-dimensional character of this superconductivity. 
From the theoretical point of view, the possibility of the spin-triplet state with the {\bf d}-vector in the CoO$_2$ plane is proposed. \cite{mochizuki,yanase} It is quite important to measure $^{17}K_c$ in the SC state using an aligned-powder sample in order to identify the SC-pairing symmetry thoroughly.

\begin{figure}[htbp]
\begin{center}
\includegraphics[width=7.5cm,clip]{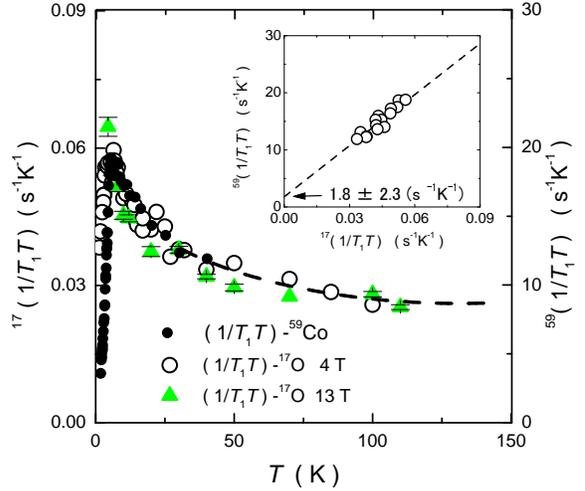}
\end{center}
\caption{$1/T_{1}T$ of $^{59}$Co and $^{17}$O. $^{59}(1/T_1T)$ is measured by a Co-NQR and $^{17}(1/T_1T)$ in 4 and 13 T. The inset shows the plot of $^{59}(1/T_{1}T)$ against $^{17}(1/T_{1}T)$. }
\label{T1T}
\end{figure}
The spin-lattice relaxation rate $1/T_{1}$ of $^{17}$O was measured at the central peak in the powder-pattern spectrum. The recovery of the nuclear magnetization after saturation pulses can be fitted by the expected theoretical curve over the entire temperature and field range. $1/T_1$ was measured in various fields, but did not show any appreciable field dependence.  
Figure \ref{T1T} shows the temperature dependence of $1/T_1T$ of $^{17}$O ($^{17}(1/T_1T)$) measured in magnetic field of 4 and 13 T, along with $1/T_1T$ of $^{59}$Co measured by Co NQR.
It was found that both results of $1/T_1T$ show the same temperature dependence. A similar result was recently reported by Ning and Imai.\cite{imai1}  We plot $^{59}(1/T_1T)$ against $^{17}(1/T_1T)$ in the inset of Fig.~\ref{T1T}. 
A good linear relationship is observed between the two quantities, indicating that $^{17}(1/T_1T)$ arises from the spin dynamics at the Co site. 
The intercept of $^{59}(1/T_1T)$ in the inset gives the relaxation rate induced by the orbital effect of the Co-$3d$ state, which is $1.8 \pm 2.3 $ s$^{-1}$ K$^{-1}$. 
It was found that the orbital contribution in $^{59}(1/T_1T)$ is negligibly smaller than the spin contribution.
The slope in the inset gives the hyperfine coupling constant at the Co site, which is estimated to be $^{59}A_{\rm hf} \sim$ 5.9 T/$\mu_B$ using $^{17}A_{\rm hf}$. 
This value is 1.2 times larger than the value estimated from the $^{59}K-\chi$ plot\cite{michioka}. 
Both results of $1/T_1T$ become enhanced below 100 K, indicative of the development of the spin fluctuations. If the $q$ dependence of the hyperfine coupling constant at the O site is taken into account, 
the AFM fluctuations at the {\bf q}-vectors far from {\bf q $=$ 0} should be excluded because such AFM fluctuations are filtered out at the O site.\cite{imai1}   

\begin{figure}[htbp]
\begin{center}
\includegraphics[width=7.5cm,clip]{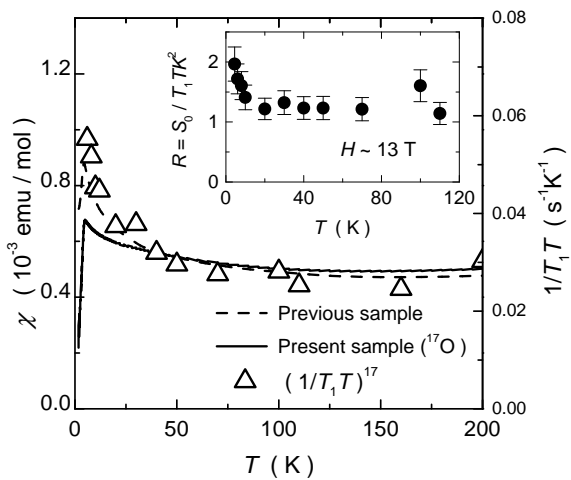}
\end{center}
\caption{Bulk susceptibilities of the present and previous samples. $1/T_{1}T$ of $^{17}$O under magnetic field of 13 T is also shown. The inset shows the temperature dependence of $R \equiv S_{0}/T_{1}TK_{\text{spin}}^{2} $(see in text).}
\label{chi}
\end{figure}
Now, we discuss the character of magnetic fluctuations in this compound on the basis of $^{17}K_{\rm spin}$ and $^{17}(1/T_1T)$.
When $K_{\rm spin}$ and $1/T_1T$ arise from noninteracting electrons, the Korringa relation, which is described as ($1/T_{1}TK^{2})_0 \equiv S_{0} = (\gamma_{\text{e}} / \gamma_{\text{n}})^{2}(\hbar/4\pi k_{\text{B}})$, holds. 
Here, $\gamma_{\rm e}$ is the gyromagnetic ratio of an electron.
The character of spin fluctuations is discussed using the value of $R$, which is estimated from a comparison between the experimental value and $S_0$ as $R \equiv ^{17}(1/T_1TK_{\rm spin}^{2})_{\rm exp} / S_{0}$.
$R$, which signifies the effect of electronic correlations, exceeds unity when antiferromagnetic (AFM) fluctuations are dominant, and it is much less than unity when ferromagnetic (FM) fluctuations become significant. 
The temperature dependence of $R$ is shown in the inset of Fig.~\ref{chi}. $R$ is temperature independent below 100 K to 20 K, and its value is $\sim 1.2$.
The value being slightly larger than unity suggests that the moderate AFM fluctuations continue down to 20 K. $R$ increases and becomes $\sim 2$ at $T_{\rm c}$, suggesting that AFM fluctuations are enhanced at low temperatures.
However, note that the magnetic fluctuations are considered to possess a critical character at low temperatures, which is suggested from the Co-NQR frequency and large $^{59}(1/T_{1}T)$ value at $T_{\rm c}$.\cite{ihara2} 
$R \sim 2$ at $T_{\rm c}$ and $\sim 1.2$ above 20 K are not so large, although the compound is situated close to the magnetic instability. Rather, if the larger Wilson ratio $R_{\rm W} \sim 2.6$ is taken into consideration, 
the presence of {\bf q $=$ 0} fluctuations is also suggested. The critical fluctuations with a moderate enhancement of $R$ may occur when the fluctuation spectrum of the AFM correlations has a peak close to {\bf q $\sim$ 0} 
and has significant contributions at {\bf q $=$ 0}. 
This is consistent with the increase in $^{17}K \propto \chi(0)$ with decreasing temperature. AFM fluctuations at {\bf q} far from {\bf q $=$ 0} would be excluded from the identical temperature dependences of $^{17}(1/T_1T)$ and $^{59}(1/T_1T)$. 
In order to reveal the entire $q$ structure in the spin-fluctuation spectrum, inelastic neutron studies of the SC compounds are highly desired.

Finally, we compare the present experimental results with the previous ones.\cite{ishida,ihara1}   
The main panel of Fig.~\ref{chi} shows the bulk susceptibilities of the present and previous samples measured at 1 T, along with $^{17}(1/T_1T)$.
In the previous paper, we suggested the presence of the FM fluctuations from the linear relationship between $1/T_1T$ and $\chi$ \cite{ishida}. 
However, the marked increase in $\chi$ below 30 K, which was observed for the previous sample, is not found for the present sample, although both samples show a gradual increase in $\chi$ in the temperature range between 150 and 30 K.
Since $\chi$ in the present sample shows a linear relationship with $^{17}K$, the moderate temperature dependence is evidenced as an intrinsic behavior, but the origin of the difference in $\chi$ below 30 K is not clear at the moment. 
In order to determine the intrinsic behavior in $\chi$ below 30 K, a systematic Knight-shift measurement of various samples is quite important. 

In conclusion, from the Knight-shift measurement at the O site in the SC state, we found a decrease in $^{17}K_{\rm spin}$ below $T_{\rm c}$ in the nonoriented powdered sample, 
which is consistent with the $^{59}$Co-NMR results \cite{kobayashi1,kobayashi2}. 
The result suggests that the superconductivity is in the spin-singlet pairing state or spin-triplet state with the SC {\bf d}-vector in the CoO$_2$ plane. $^{17}K_{\rm spin}$ is evaluated from the $^{17}K-\chi$ plot, 
and is compared with $^{17}(1/T_1T)$. The development of AFM fluctuations is suggested below 30 K, and a clear indication of FM fluctuations was not observed in the $^{17}$O NMR measurements. However, 
from the larger Wilson Ratio $R_{\rm W} \sim 2.6$ and the identical temperature dependences of $^{17}(1/T_1T)$ and $^{59}(1/T_1T)$, incommensurate fluctuations with small wave vectors are suggested to be enhanced at low temperatures, 
which have a significant contribution at {\bf q $=$ 0}.

We thank C. Michioka and Y.~Maeno for experimental support and valuable discussions. We also thank H.~Ikeda, S.~Fujimoto, K.~Yamada, Y.~Yanase, M. Mochizuki, and M.~Ogata for valuable discussions.
This work was partially supported by CREST of the Japan Science and Technology Corporation (JST) and the 21 COE program on ``Center for Diversity and Universality in Physics'' from MEXT of Japan, and by Grants-in-Aid for Scientific Research from the Japan Society for the Promotion of Science (JSPS) and MEXT.


\begin{thebibliography}{99}

\bibitem{takada}
K.~Takada, H.~Sakurai, E.~Takayama-Muromachi, F.~Izumi, R.~A.~Dilanian and T.~Sasaki: Nature {\bf 422} (2003) 53.

\bibitem{ishida}
K.~Ishida, Y.~Ihara, Y.~Maeno, C.~Michioka, M.~Kato, K.~Yoshimura, K.~Takada, T.~Sasaki, H.~Sakurai and E.~Takayama-Muromachi: JPSJ {\bf 72} (2003) 3041.

\bibitem{fujimoto}
T.~Fujimoto, G-q.~Zheng, Y.~Kitaoka, R.~L.~Meng, J.~Cmaidalka and C.W.~Chu: PRL {\bf 92} (2004) 047004.

\bibitem{yang}
H.D.~Yang, J.-Y.~Lin, C.P.~Sun, Y.C.~Kang, C.L.~Huang, K.~Takada, T.~Sasaki, H.~Sakurai and E.~Takayama-Muromachi: PR{\bf B} {\bf 71} (2005) 020505.

\bibitem{ihara1}
Y.~Ihara, K.~Ishida, C.~Michioka, M.~Kato, K.~Yoshimura, K.~Takada, T.~Sasaki, H.~Sakurai and E.~Takayama-Muromachi: JPSJ {\bf 73} (2004) 2069.

\bibitem{ihara2}
Y.~Ihara, K.~Ishida, C.~Michioka, M.~Kato, K.~Yoshimura, K.~Takada, T.~Sasaki, H.~Sakurai and E.~Takayama-Muromachi: JPSJ {\bf 74} (2005) 867.

\bibitem{mochizuki}
M.~Mochizuki, Y.~Yanase and M. Ogata: PRL {\bf 94} (2005) 147005.

\bibitem{Ikeda}
H.~Ikeda, and Y.~Nisikawa and K.~Yamada: JPSJ {\bf 73} (2005) 17. 

\bibitem{yanase}
Y.~Yanase, M.~Mochizuki and M.~Ogata: JPSJ {\bf 74} (2005) 430.

\bibitem{kobayashi1}
Y.~Kobayashi, M.~Yokoi and M.~Sato: JPSJ {\bf 72} (2003) 2453.

\bibitem{kobayashi2}
Y.~Kobayashi, H.~Watanabe, M.~Yokoi, T.~Moyoshi, Y.~Mori and M.~Sato: cond-mat/0412466.

\bibitem{waki}
T.~Waki, C.~Michioka, M.~Kato, K.~Yoshimura, K.~Takada, H.~Sakurai, E.~Takayama-Muromachi and T.~Sasaki: cond-mat/0306036.

\bibitem{sakurai1}
H.~Sakurai {\it et al.}, private communication.

\bibitem{MetallicShift}
G.~C.~Carter, L.H.Bennett, and D.J.Kahan, ``Metallic Shift in NMR'' (Pergamon Press, Oxford OX30BW, England,)  

\bibitem{sakurai2}
H.~Sakurai, K.~Takada, S.~Yoshii, T.~Sasaki, K.~Kindo and E.~Takayama-Muromachi: PR{\bf B} {\bf 68} (2003) 132507.   

\bibitem{chou}
F.C.~Chou, J.H.~Cho, P.A.~Lee, E.T.~Abel, K.~Matan and Y.S.~Lee: PRL {\bf 92} (2004) 157004.

\bibitem{ishida2}
K.~Ishida {\it et al.}: JPSJ {\bf 62} (1993) 2803.

\bibitem{imai1}
F.L.~Ning and T.~Imai: PRL {\bf 94} (2005) 227004.

\bibitem{michioka}
C. Michioka {\it et al.}: cond-mat/0403293.


\end{thebibliography}
\end{document}